\documentclass[twocolumn,floatfix,prb,aps,showpacs]{revtex4-1}
\usepackage{graphicx,amsmath,amssymb}

\begin{document}

\title{Intra-Landau level magnetoexcitons and the transition between quantum Hall states in undoped bilayer graphene}
\author{Csaba T\H oke$^{1,2}$ and Vladimir I. Fal'ko$^{2}$}
\affiliation{$^{1}$Institute of Physics, University of P\'ecs, 7624 P\'ecs, Hungary}
\affiliation{$^{2}$Department of Physics, Lancaster University, Lancaster, LA1 4YB, United Kingdom}
\date{\today}

\begin{abstract}
We study the collective modes of the quantum Hall states in undoped bilayer graphene in a strong perpendicular magnetic and electric field.
Both for the well-known ferromagnetic state that is relevant for small electric field $E_\perp$ and the analogous layer polarized one suitable for large $E_\perp$,
the low-energy physics is dominated by magnetoexcitons with zero angular momentum that are even combinations of excitons that conserve Landau orbitals.
We identify a long wave length instability in both states, and argue that there is an intermediate range of the electric field
$E^{(1)}_\text{c} <E_\perp<E^{(2)}_\text{c}$ where a gapless phase interpolates between the incompressible quantum Hall states.
The experimental relevance of this crossover via a gapless state is discussed.
\end{abstract}


\maketitle

\section{Introduction}

\begin{figure}[htbp]
\begin{center}
\includegraphics[width=0.75\columnwidth, keepaspectratio]{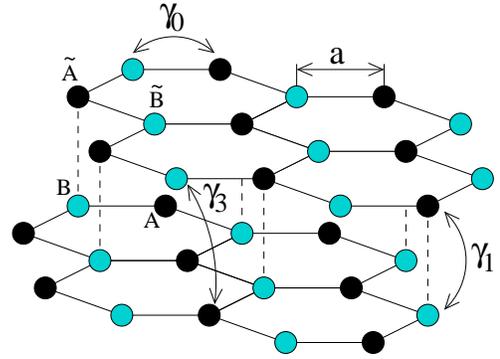}
\end{center}
\caption{\label{intro}
(Color online)
The crystalline structure of bilayer graphene in Bernal stacking.
$A,B$ ($\widetilde A,\widetilde B$) denote inequivalent sites on two hexagonal carbon sublattices on the bottom (top) layer.
Hopping amplitudes are indicated.
}
\end{figure}

Recent magnetotransport experiments \cite{Feldman,Zhao,Weitz,Kim} on bilayer graphene \cite{Novoselov} crystal in Bernal stacking
(Fig.~\ref{intro}) have shown that its characteristic, eight-fold quasi-degenerate (spin $\sigma=\uparrow,\downarrow$, valley $\xi=\pm1$, and Landau orbital $n=0,1$)
central Landau band (CLB) is split at all integer values of the filling factor $\nu=\rho h/eB_\perp$.
Quantum Hall ferromagnetic states have been suggested \cite{Barlas} to explain these states.
At $\nu=0$, on the other hand, the longitudinal resistivity $\rho_{xx}$ diverges beyond a sample-dependent threshold value of the magnetic field both for
decreasing temperature and increasing field. Similar behavior has been observed in monolayers \cite{slg} and in particle-hole symmetric semiconductor systems.\cite{Gusev}
While this is unusual, it does not rule out quantum Hall physics, because Laughlin's gauge argument \cite{gauge}
connects a vanishing longitudinal conductivity $\sigma_{xx}$ to the quantized Hall conductivity $\sigma_{xy}$,
which is consistent \cite{DasSarma} with a divergent $\rho_{xx}=\rho_{yy}$ for $\rho_{xy}=0$.
While Zhao \textit{et al.} \cite{Zhao} found that the gap at $\nu=0$ depends on the field as $\sqrt{B_\perp}$, as expected of Coulomb interaction effects,
Feldman \textit{et al.},\cite{Feldman} using high-mobility suspended samples, measured a gap that opens linearly with $B_\perp$ and hardly depends on $B_\parallel$.
The latter finding suggests that many-body effects dominate the Zeeman splitting.
The observed linear $B_\perp$ dependence of the gap up to a rather high threshold value has been explained by Nandkishore and Levitov \cite{Nand} and Gorbar \textit{et al.} \cite{Gorbar} taking dynamical screening into account.
Thermally activated transport across the $\propto B_\perp$ gap then explains \cite{Nand,Gorbar} the exponential growth of resistivity $\rho_{xx}$ in $B_\perp/k_B T$ found in Ref.~\onlinecite{Feldman}.

At zero magnetic field, recent experiments \cite{Zhang} have confirmed the emergence of a band gap \cite{McCann} if a perpendicular electric field is applied.
In the presence of a perpendicular magnetic field $B_\perp$, the band gap affects the Landau levels (LL's); but as exchange energy considerations are fundamental in the quantum Hall regime,
the low-energy physics at integer filling factors is determined by the \textit{interplay} of the electron-electron interaction, the Zeeman energy $\Delta_Z=g\mu_BB_\perp$,
and the interlayer potential energy difference $\Delta_L=edE_\perp-\Delta n/(2\epsilon_0)$ between the layers
($\Delta n$ is the electron density imbalance, $d=0.335$ nm is the distance between the layers, $g$ is the gyromagnetic factor, and $\mu_B$ is the Bohr magneton).
In a mean-field approximation, the ground-state of this system spontaneously breaks \cite{McCann,Gorbar} the (approximate) spin and valley symmetry.\cite{Barlas}
Obeying variants of Hund's rule, the ground-state is ferromagnetic in the $\Delta_L\ll\Delta_Z$ limit and valley-polarized in the $\Delta_L\gg\Delta_Z$ limit.
As the Coulomb interaction, however, is quite strong in comparison to the LL splitting, $[e^2/(4\pi\epsilon_0\epsilon_r\ell_B)]/E_2\approx 11.7/[\sqrt{B\;(\text{T})}\epsilon_r]$ ($E_2$ is the greatest
Landau level energy difference, c.f.\ Sec.~\ref{secll} below), screening by inter-LL transitions might be important. \cite{Nand,Gorbar}
(Here $\ell_B=\sqrt{\hbar/eB_\perp}$ is the magnetic length, and $\epsilon_r$ is the relative dielectric constant of the bilayer graphene sheet and its immediate environment.)
Thus, when studying the excitations of bilayer graphene on the $\sigma_{xy}=0$ plateau, we may start from these symmetry-breaking quantum Hall states.
We shall regard, however, the perpendicular electric field $E_\perp$ instead of $\Delta_L$ as the tunable parameter.

The paper is organized as follows. In Sec.~\ref{secll}, we review the Landau-level structure of bilayers and recall the symmetry breaking ground-states that are relevant at charge neutrality.
In Sec.~\ref{secmagnon}, we define the type of excitations we study.
In Sec.~\ref{secresults}, our results are presented, and in Sec.~\ref{secdiscussion}, their significance is discussed.
Technical details are included in the Appendixes.

\section{Landau levels and ground-states}
\label{secll}

In the vicinity of the valley centers corresponding to the $K$ ($\xi=1$) and $K'$ ($\xi=-1$) first Brillouin zone corners, the electronic structure of bilayer graphene is well described by the tight-binding Hamiltonian \cite{tightbinding}
\begin{equation}
\label{hamilton}
\hat H_\xi=\xi\begin{pmatrix}
\frac{\Delta_L}{2} &  0 & 0 &  v\pi^\dag \\
0 & -\frac{\Delta_L}{2} &  v\pi & 0 \\
0 &  v\pi^\dag & -\frac{\Delta_L}{2} & \xi\gamma_1 \\
v\pi & 0 & \xi\gamma_1 & \frac{\Delta_L}{2}
\end{pmatrix} -\Delta_Z\hat\sigma_z,
\end{equation}
where  $\pi=p_x+ip_y$ and $\mathbf p=-i\hbar\nabla -e\mathbf A$,
$v=\sqrt3 a\gamma_0/2\hbar\approx 10^6$ m/s is the intra-layer velocity, and $\gamma_1\approx0.39$ eV is the inter-layer hopping amplitude.
This Hamiltonian acts in the basis of sublattice Bloch states $[\psi_A,\psi_{\widetilde B},\psi_{\widetilde A},\psi_B]$ in valley $K$ and $[\psi_{\widetilde B},\psi_A,\psi_B,\psi_{\widetilde A}]$ in valley $K'$.
Having a suspended sample in mind, we will assume $\epsilon_r=1,g=2$.

We emphasize that our main results follows from the four-band model of Eq.~(\ref{hamilton}); the low-energy two-band model\cite{McCann} is only occassionally referred to for contrast.
We have neglected $\gamma_3,\gamma_4,\Delta'$ of the Slonczewski-Weiss-McClure model \cite{SWM} as usual.
In particular, Ref.~\onlinecite{tightbinding} has shown that $\gamma_3\approx0.1\gamma_0$ has no significant effect on the Landau levels of bilayer graphene, which suggests that orbital mixing due to $\gamma_3$ is negligible.
Using typical values from the literature, $\gamma_4\approx40$ meV and $\Delta'\approx7$ meV are small in comparison to the Coulomb energy scale $e^2/(4\pi\epsilon_0\epsilon_r\ell_B)=56\text{ meV} \sqrt{B[\text{T}]}/\epsilon_r$.

Using the gauge $\mathbf A=-By\mathbf{\hat x}$, the Landau levels and orbitals are obtained from the ansatz \cite{Pereira,magnetoelectric}
\begin{gather}
\psi_{0q}=\left[\eta_{0q}, 0, 0, 0\right]^{\text{T}},\label{states}\\
\psi_{1\alpha q}=\left[A^1_{\alpha\xi}\eta_{1q}, 0, iC^1_{\alpha\xi}\eta_{0q}, iD^1_{\alpha\xi}\eta_{0q}\right]^{\text{T}},\nonumber\\
\psi_{n\alpha q}=\left[A^n_{\alpha\xi}\eta_{nq}, B^n_{\alpha\xi}\eta_{n-2,q}, iC^n_{\alpha\xi}\eta_{n-1,q}, iD^n_{\alpha\xi}\eta_{n-1,q}\right]^{\text{T}},\nonumber
\end{gather}
where $n\ge2$; $1\le\alpha\le4$ ($E_{n\alpha}<E_{n,\alpha+1}$), and $\eta_{nq}$ are the single-particle states in the ordinary two-dimensional electron gas with quadratic carrier dispersion,
\begin{equation}
\label{etaeq}
\eta_{nq}(\mathbf r)=
\frac{e^{iqx-\left(y/\ell_B-q\ell_B\right)^2/2}}{\sqrt{2\pi\sqrt\pi 2^n n!\ell_B}}H_n\left( \frac{y}{\ell_B}-q\ell_B \right),
\end{equation}
and $H_n$ is a Hermite-polynomial.
Introducing the dimensionless quantities
\[
t=\gamma_1\ell_B/v\hbar,\quad\delta=\Delta_L\ell_B/2 v\hbar,
\]
the $n=1$ Landau levels $E_{1\alpha\xi}=\hbar v\epsilon_{1\alpha\xi}/\ell_B$, $\alpha=1,2,3$, are obtained from the ensuing secular equation, and the orbitals are determined by
\[
A^1_{\alpha\xi}=-\frac{\xi\left(t^2-\epsilon^2+\delta^2\right)}{\sqrt 2 tN_{1\alpha}},\quad
C^1_{\alpha\xi}=\frac{1}{N_{1\alpha}},\quad D^1_{\alpha\xi}=\frac{\epsilon+\xi\delta}{tN_{1\alpha}},
\]
where $N_{1\alpha}$ is the appropriate normalization factor.
The $n\ge1$ Landau levels $E_{n\alpha\xi}=\hbar v\epsilon_{n\alpha\xi}/\ell_B$, $\alpha=1,2,3,4$ follow similarly,
The orbitals are specified by
\begin{align*}
A^n_{\alpha\xi}&=\frac{1}{N_{n\alpha}},\quad
D^n_{\alpha\xi}=-\frac{\xi(\epsilon-\xi\delta)}{\sqrt{2n}N_{n\alpha}},\\
C^n_{\alpha\xi}&=\frac{\xi}{tN_{n\alpha}}\left(\sqrt{2n}-\frac{(\epsilon-\xi\delta)^2}{\sqrt{2n}}\right),\\
B^n_{\alpha\xi}&=\frac{\xi\sqrt{2n-2}}{(\epsilon+\xi\delta)N_{n\alpha}}C^n_{\alpha\xi},
\end{align*}
with the appropriate normalization factor  $N_{n\alpha}$.

For the central Landau bands, in particular, we obtain
\begin{gather*}
E_{1,\alpha=2,\xi\sigma}=\frac{\xi\Delta_L}{2}
\left(\frac{t^2-2}{t^2+2}+\frac{32\delta^2t^2}{(t^2+2)^4} + O(\delta^5)\right)+\Delta_Z\sigma_z,\notag\\
E_{0\xi\sigma}=\frac{\xi\Delta_L}{2}+\Delta_Z\sigma_z.
\end{gather*}
States $\psi_{0\xi\sigma q}$ and $\psi_{1,\alpha=2,\xi\sigma q}$ form the central Landau band (CLB) octet.
For $n=0$ ($n=1$) the states are completely (predominantly) located on $A$ sites in valley $K$ and $\widetilde B$ sites in valley $K'$,
making valley equivalent to layer in the central Landau band.

With this notation, the ferromagnetic state is\cite{Barlas}
$|\text{f.m.}\rangle=\prod_{q\xi}\hat a^\dag_{0,\xi,\uparrow q}\hat a^\dag_{1,\alpha=2,\xi,\uparrow q}|0\rangle$,
and the valley (layer) polarized one is
$|\text{v.p.}\rangle=\prod_{q\sigma}\hat a^\dag_{0,\xi=1,\sigma q}\hat a^\dag_{1,\alpha=2,\xi=1,\sigma q}|0\rangle$.
Here $|0\rangle$ is the ``vacuum'' of an infinite number of filled valence Landau bands ($n\ge2$, $\alpha=2,3$, all combinations of $\xi=\pm1$ and $\sigma=\uparrow,\downarrow$), and
$\hat a^\dag_{n\alpha\xi\sigma p}\equiv \hat a^\dag_{Np}$ creates a particle in state $\psi_{Np}$.
$\Delta_L$ differs in the two states:
\begin{equation}
\label{screen}
\Delta_L^{\text{f.m.}} = eE_\perp d,\quad\quad
\Delta_L^{\text{v.p.}} = ed\left(E_\perp - E_0\right),
\end{equation}
where $E_0\equiv e^2 d/\left(\pi\epsilon_0\epsilon_r\ell_B^2\right)\approx8.75\times10^6 B\text{[T]}/\epsilon_r$ V/m.
Thus in state $|\text{v.p.}\rangle$ $\Delta_L$ is reduced from its unscreened value by $\sim 34 B\text{[T]}/\epsilon_r$ K,
which is greater than the Zeeman energy $\Delta_Z\approx 1.4B\text{[T]}$ K.

\section{Magnetoexcitons}
\label{secmagnon}

The low-energy excitations of quantum Hall states are magnetoexcitons, which are obtained by promoting an electron from a filled Landau band to an empty band \cite{Kallin}.
These neutral excitations have a well-defined center-of-mass momentum $\mathbf Q$.
They are collective, but also include widely separated particle-hole pairs in the $Q\to\infty$ limit.
The latter limit determines the transport gap unless skyrmions form \cite{skyrmions}.
Magnetoexcitons are created from the ground-state by operators \cite{Falko} $\hat\Psi^\dag_{NN'}(\mathbf Q)$,
\begin{equation}
\label{exciton}
\hat\Psi^\dag_{NN'}(\mathbf Q)=\sum_p e^{ipQ_y\ell_B^2}\hat a^\dag_{Np}\hat a_{N'p-Q_x}.
\end{equation}

In the low magnetoexciton density limit, the interaction between magnetoexcitons is neglected; then it is sufficient to diagonalize the mean-field Hamiltonian matrix
\begin{multline}
H^{(\tilde N\tilde N')}_{(NN')}(\mathbf Q)=\langle 0|\Psi_{\tilde N\tilde N'}(\mathbf Q)\hat V\Psi^\dagger_{NN'}(\mathbf Q) |0\rangle\\
-\delta_{N\tilde N}\delta_{N'\tilde N'}\langle0|\hat V|0\rangle,
\label{hamirest}
\end{multline}
where $N=(n,\alpha,\xi,\sigma)$ [$N'=(n',\alpha',\xi',\sigma')$] specifies the Landau band where the particle (hole) is created;
$\delta_{NN'}=\delta_{\sigma\sigma'}\delta_{\xi\xi'}\delta_{nn'}\delta_{\alpha\alpha'}$.
Taking the four-spinor structure of the Landau orbitals is the only nonstandard step. Details are delegated to the Appendixes.

Magnetoexcitons carry spin and pseudospin (valley) quantum numbers, as defined by the particle and hole Landau bands involved.
In addition, an ``angular momentum quantum number'' $l_z=|n|-|n'|$ can be defined as usual.\cite{Kallin,Iyengar}
This quantity, however, is exactly conserved only in the $\mathbf Q\to0$ limit.

\section{Results}
\label{secresults}

\begin{figure}[!htb]
\begin{center}
\includegraphics[width=\columnwidth, keepaspectratio]{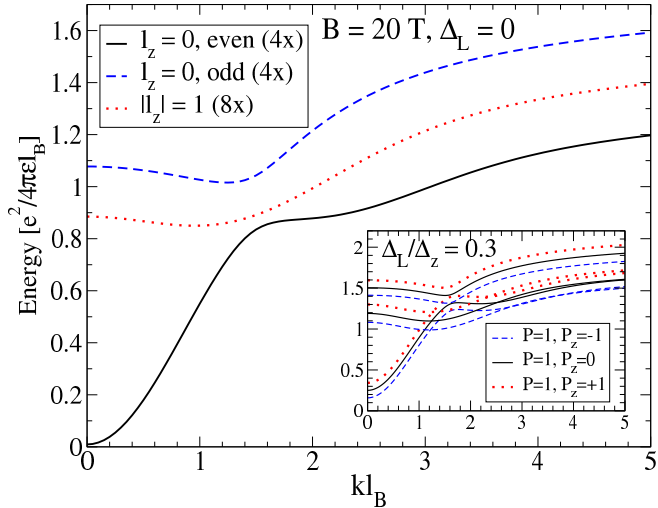}
\end{center}
\caption{\label{dispersion4} (Color online)
Excitations of the neutral graphene bilayer in perpendicular fields in the absence of an interlayer energy difference $\Delta_L$.
Even and odd refer to the modes $E^{P,P_z,\text{even}}_{l_z=0}(\mathbf k)$ and $E^{P,P_z,\text{odd}}_{l_z=0}(\mathbf k)$,
which at low momenta tend to the even and odd linear combinations, $\propto{\Psi^\dag}^{P,P_z}_{(0\xi\downarrow,0\xi'\uparrow)}\pm{\Psi^\dag}^{P,P_z}_{(1\xi\downarrow,1\xi'\uparrow)}$, of
excitations that conserve Landau orbitals, respectively.
The inset illustrates qualitatively the band splittings by $\Delta_L$ for unrealistic Zeeman energy ($B\approx10^4$ T) for visibility.
}
\end{figure}

\begin{figure}[!htb]
\begin{center}
\includegraphics[width=\columnwidth, keepaspectratio]{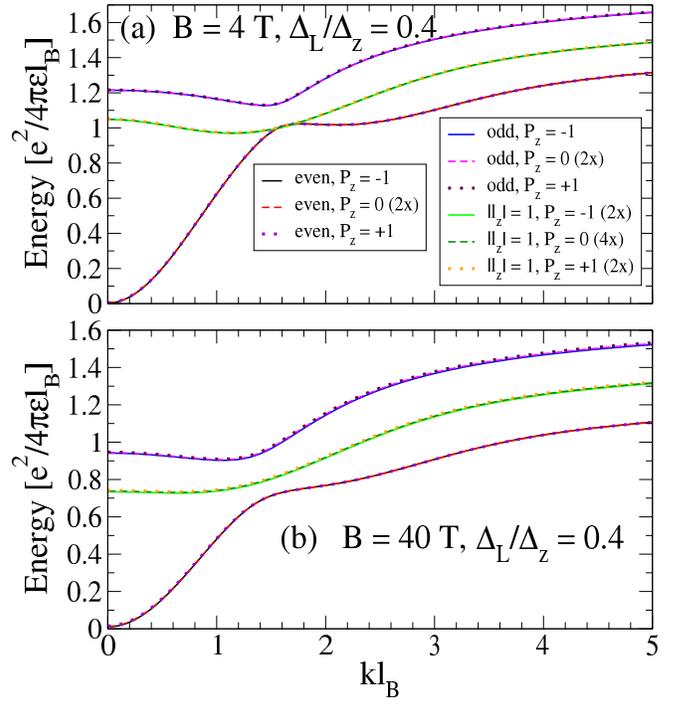}
\end{center}
\caption{\label{dispersionB} (Color online)
Excitations of the neutral graphene bilayer in perpendicular fields at various parameters $B_\perp$ and $\Delta_{\text{L}}/\Delta_{\text{Z}}$ as indicated on the panels.
The energy unit $e^2/(4\pi\epsilon_0\epsilon_r\ell_B)$ already scales with $B$; the additional change shown here is due to the $B$ dependence of the Landau orbitals.
}
\end{figure}

The magnetoexcitons of the ferromagnetic state $|\text{f.m.}\rangle$ are characterized by their pseudospin $P,P_z$, and angular momentum $l_z$ quantum numbers.
The latter is $l_z=1$ when an electron is promoted from an $n=0$ orbital to $n=1$, and $l_z=-1$ in the converse transition.
(For $|\text{v.p.}\rangle$, spin and pseudospin are interchanged.)
The spectra are shown in Fig.~\ref{dispersion4} for $\Delta_L=0$.
The $l_z=0$ excitations have two branches, $E^{P,P_z,\text{even}}_{l_z=0}(\mathbf k)$ and $E^{P,P_z,\text{odd}}_{l_z=0}(\mathbf k)$, each with a pseudospin triplet and a pseudospin singlet band.
In the long-wavelength limit, the $E^{P,P_z,\text{even}}_{l_z=0}(\mathbf k)$ mode tends to $\sim\Delta_Z+P_z\Delta_L$ (apart from a small $B_\perp$-dependent correction).
This excitation is $\propto{\Psi^\dag}^{P,P_z}_{(0\xi\downarrow,0\xi'\uparrow)}+{\Psi^\dag}^{P,P_z}_{(1\xi\downarrow,1\xi'\uparrow)}$ in this limit,
while the gapped $E^{P,P_z,\text{odd}}_{l_z=0}(\mathbf k)$ mode is $\propto{\Psi^\dag}^{P,P_z}_{(0\xi\downarrow,0\xi'\uparrow)}-{\Psi^\dag}^{P,P_z}_{(1\xi\downarrow,1\xi'\uparrow)}$.
The $l_z=\pm1$ branches are gapped and degenerate, in compliance with the global particle-hole symmetry of the system.

\begin{figure}[!htb]
\begin{center}
\includegraphics[width=\columnwidth, keepaspectratio]{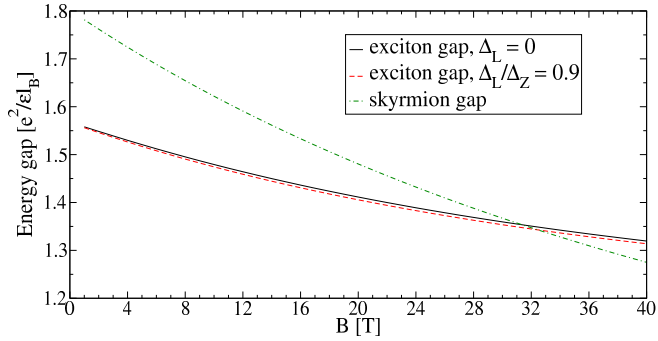}
\end{center}
\caption{\label{gap} (Color online)
Charge gap from widely separated particle hole pairs (excitons) and skyrmion-antiskyrmion pairs.
The latter is independent of the potential difference between layers $\Delta_L$; the former is shown for two different values of $\Delta_L$.
}
\end{figure}

For nonzero $\Delta_L$, each band is split according to its pseudospin and spin quantum numbers as sketched in the inset\cite{fnote} of Fig.~\ref{dispersion4}.
For generic $\Delta_L$, the dispersion of magnetoexcitons also depends directly on $B_\perp$ (not just through $\ell_B$) because of the $\delta=\Delta_L\ell_B/2 v\hbar$ dependence of the orbitals; c.f.\ Fig.~\ref{dispersionB}.
The excitation energies are reduced by roughly 1\% to 16\% from the two-band value $\frac{5}{4}\sqrt{\frac{\pi}{2}}e^2/(4\pi\epsilon_0\epsilon_r\ell_B)$, roughly linearly, in the range $1$ T $ \le B_\perp\le 40$ T.
The density-of-states peak due to the flat region in the upper $l_z=0$ branch and the $l_z=\pm1$ branches might be observable in Raman scattering.\cite{Pinczuk}
The lower $l_z=0$ branch is less curved for increasing $B_\perp$; but as Fig.~\ref{gap} demonstrates,
this reduced spin stiffness is insufficient to make the charge gap skyrmionic unless, perhaps, for huge fields $B_\perp>30$ T.\cite{double}

\begin{figure}[!htb]
\begin{center}
\includegraphics[width=\columnwidth, keepaspectratio]{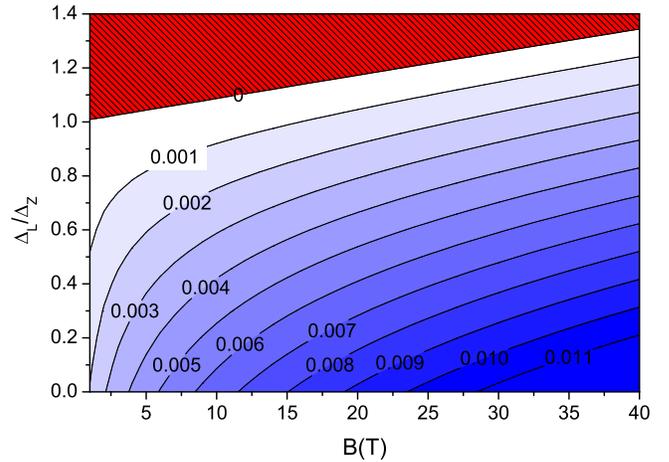}
\end{center}
\caption{\label{phase} (Color online)
Zero temperature phase diagram of undoped bilayer graphene.
The red shaded region above the thick line corresponds to a compressible state.
The contours in the ferromagnetic phase show the long-wavelength limit of the magnetoexciton gap.
}
\end{figure}

Starting in the $\Delta_L\ll\Delta_Z$ limit in the ferromagnetic state, there is a long-wavelength instability due to the pseudospin triplet mode,
\begin{multline}
{\Psi^\dag}^{P=1,P_z=-1,\text{even}}_{l_z=0}(\mathbf k)\approx\\
\frac{1}{\sqrt2}\left(
\Psi^\dag_{(0\xi=-1,\uparrow,0\xi'=1,\downarrow)}+\Psi^\dag_{(1\xi=-1,\uparrow,1\xi'=1\downarrow)}
\right)
\label{cmode}
\end{multline}
at $\Delta_Z\sim\Delta_L$, where the excitation energy of the mode (see Fig.~\ref{dispersion4}) reaches zero.
This is a signal of a transition to the other state, which becomes energetically favored at this point [the strict equality in Eq.~(\ref{cmode}) holds at $\mathbf k=0$].
Figure \ref{phase} shows the regions of stability in terms of $B_\perp$ and $\Delta_L/\Delta_Z$.
Conversely, starting from the valley-polarized state in the  ($\Delta_L\gg\Delta_Z$) limit, there is an instability due to the mode
\begin{multline}
{\Psi^\dag}^{S=1,S_z=-1,\text{even}}_{l_z=0}(\mathbf k)\approx\\
\frac{1}{\sqrt2}\left(
\Psi^\dag_{(0\xi=1,\downarrow,0\xi'=-1,\uparrow)}+\Psi^\dag_{(1\xi=1,\downarrow,1\xi'=-1\uparrow)}
\right)
\label{cmode2}
\end{multline}
For realistics fields, the phase boundary is given by (Fig.~\ref{phase})\cite{compare}
\begin{equation}
\label{boundary}
\frac{\Delta_L}{\Delta_Z}\approx1+0.00875 B[\text{T}].
\end{equation}

\section{Discussion}
\label{secdiscussion}

In experiments, the energy difference between the layers, $\Delta_L$, is determined indirectly via the electric field $E_\perp$.
By Eq.~(\ref{screen}) there is a region
$E_{\text{c}}^{(1)}<E_\perp<E_{\text{c}}^{(2)}$, with 
$E_{\text{c}}^{(1)}=\Delta_Z/ed\approx 3.5\times10^5B{[T]}$ V/m and 
$E_{\text{c}}^{(2)}=\Delta_Z/ed+ e/(\pi\epsilon_0\ell_B^2)\approx 9.1\times10^6B{[T]}$ V/m,
where the ferromagnetic state is destabilized by the lowest $l_z=0$ mode,
while the valley-polarized state is not yet stable (it would overscreen the external electric field and $\Delta_L>\Delta_Z$ would not hold).
For $E_\perp>E_{\text{c}}^{(2)}$, the valley-polarized state becomes the ground-state.

As $|\text{v.p.}\rangle$ and $|\text{f.m.}\rangle$ have extremal total Zeeman and layer polarization energy,
there is no energetically more favorable quantum Hall ferromagnet in the $E_{\text{c}}^{(1)}<E_\perp<E_{\text{c}}^{(2)}$ range.
When $E_\perp$ reaches $E^{(1)}_{\text{c}}$, the ground-state energy can be lowered by creating magnetoexcitons.
Two scenarios are consistent with this observation: (i) a first-order phase transition between the ferromagnetic and the valley-polarized quantum Hall states, and
(ii) a crossover via compressible states in the $E_{\text{c}}^{(1)}<E_\perp<E_{\text{c}}^{(2)}$ range.
At zero temperature, this issue is decided by the energy comparison of the Maxwell construction that interpolates between the energies of the two states $|\text{f.m.}\rangle$ and $|\text{v.p.}\rangle$,
and the state that is obtained from these quantum Hall states by populating the $\mathbf k=0$ state of magnetoexcitons of the critical mode of Eq.~(\ref{cmode}).
Formally, the latter can be conceived as a Bose-Einstein condensation of magnetoexcitons, but it merely amounts to a gradual change of the many-body ground-state in the crossover, with no quasiparticles keeping their identity while condensed.
If the magnon-magnon interaction is neglected, these energies are degenerate.
If, on the other hand, the interaction between magnetoexcitons assumes a van der Waals profile, as is typical for the inter-exciton interaction in analogous systems at vanishing magnetic field,\cite{Moska}
the energy of the compressible crossover state is lower than that of the phase-separated quantum Hall systems, thus, at least in the neighborhood of $E_{\text{c}}^{(1)}$ and $E_{\text{c}}^{(2)}$ where the
two-magnon interaction dominates over many-magnon effective interaction terms, the gapless state prevails.
The calculation of the magnon-magnon interaction is delegated to future work.

Most experiments have been performed \cite{Feldman,Zhao} with a single backgate.
The gate voltage to counter possible extrinsic doping was $|V_D|<0.5$ V in Ref.~\onlinecite{Feldman} and $V_D\approx4$ V in Ref~\onlinecite{Zhao}.
The electric field then must be a few volts on about 300 nm, i.e., $\sim3\times 10^{6}$ V/m.
With $B\sim10$ T, this must be within the ferromagnetic region.
Moreover, the $g$-factor might be significantly enhanced \cite{magnetoelectric}.
Thus we expect that only dual-gated bilayer graphene \cite{dualgate,Weitz} can be tuned into the gapless state.

More recently, dual-gated two-terminal magnetotransport measurements were reported by Wietz \textit{et al.} \cite{Weitz}.
A transition exhibiting increased conductance was observed between two $\sigma\approx0$ plateaus in the two-terminal conductance as a function
of the external electric field for not too small magnetic fields.
The threshold field $E_\perp$ of the transition is about 40\% greater than our $E^{(2)}_c$, and its slope as a function of $B_\perp$, 11 mV/(nm$\cdot$T), compares well with our prediction of 9 mV/(nm$\cdot$T).
Admittedly, the calculation we present would imply a wider conducting region than that found in Ref.~\onlinecite{Weitz}.
Four-terminal experiments by Kim, Lee, and Tutuc\cite{Kim} also found a linear dependence of the critical electric field with a slightly higher slope, 12 to 18 mV/(nm$\cdot$T), and a more stable ferromagnetic state.
Inter-Landau level screening, neglected in this paper, may stabilize the ferromagnetic state, in accordance with Refs.~\onlinecite{Gorbar,Nand}.
A more conclusive comparison between theory and experiment requires further progress.

To summarize, we have presented the excitation spectra of undoped bilayer graphene in the quantum Hall regime when the interlayer bias is introduced by a perpendicular electric field.
The collective modes of zero angular momentum that are an even combination of transitions that conserve the Landau orbital give rise to a long-wavelength instability,
which precipitates a compressible region between two quantum Hall states for a range of the perpendicular electric field.

\begin{acknowledgements}
This work was supported by the Lancaster University-EPSRC Portfolio Partnership.
C. T. was supported by Science, Please! Innovative Research Teams, SROP-4.2.2/08/1/2008-0011.
We thank Judit S\'ari for carefully reading the manuscript. 
\end{acknowledgements}

\appendix

\section{Mean-field theory of magnetoexcitons in bilayer graphene}

While the mean-field theory of magnetoexcitons is standard material, we present its adaptation to bilayer graphene.
The mean-field approach amounts to the diagonalization of  the Hamiltonian matrix $H^{(\tilde N\tilde N')}_{(NN')}(\mathbf Q)$ in Eq.~(\ref{hamirest}), where $N\equiv(n,\alpha,\xi,\sigma)$ specifies a Landau band.
Suppressing spin and valley labels for simplicity, the field operators for bilayer graphene Landau orbitals are
\begin{multline}
\label{fieldop}
\hat\Psi(\mathbf r)=\sum_{n\alpha\xi\sigma q}\left[
A^n_{\alpha\xi}\eta_{nq}(\mathbf r), B^n_{\alpha\xi}\eta_{n-2,q}(\mathbf r),\right.\\
\left.iC^n_{\alpha\xi}\eta_{n-1,q}(\mathbf r),iD^n_{\alpha\xi}\eta_{n-1,q}(\mathbf r)\right]^{\text{T}}\hat a_{n\alpha\xi\sigma q},
\end{multline}
cf.\ Eqs.~(\ref{etaeq}) and (\ref{states}), and $A^n_{\alpha\xi},B^n_{\alpha\xi},C^n_{\alpha\xi},D^n_{\alpha\xi}$ are set to zero whenever redundant for $n=0,1$.
(That is, we set $A^0_{\alpha\xi}=0$ for $\alpha=2,3,4$, $B^0_{\alpha\xi}=C^0_{\alpha\xi}=D^0_{\alpha\xi}=B^1_{\alpha\xi}=0$ for $\alpha=1,2,3,4$, and $A^1_{4}=B^1_{4}=C^1_{4}=D^1_{4}=0$.)
The Coulomb Hamiltonian is
\begin{multline}
\label{hamiorigin}
\hat H=\frac{1}{2}\int d\mathbf xd\mathbf x' \hat\Psi^\dag(\mathbf x)\otimes\Psi^\dag(\mathbf x')\hat V(\mathbf x-\mathbf x')\Psi(\mathbf x')\otimes\Psi(\mathbf x),
\end{multline}
where $\hat V$ has a matrix structure as it is sandwiched between spinors.
If $\beta$ runs over spinorial indices ($\beta=1,2,3,4$ for sites $A,\widetilde B,\widetilde A,B$ in valley $K$ and $\widetilde B,A,B,\widetilde A$ in valley $K'$, respectively), $\hat V$ has the form
\begin{multline}
\label{interaction}
V_{\xi_1\beta_1\xi_2\beta_2,\xi'_2\beta'_2\xi'_1\beta'_1}= \frac{e^2\delta_{\beta_1\beta'_1}\delta_{\beta_2\beta'_2}\delta_{\xi_1\xi'_1}\delta_{\xi_2\xi'_2}}{4\pi\epsilon_0\epsilon_r}\\
\left[ \frac{\delta_{L(\xi_1,\beta_1),L(\xi_2,\beta_2)}}{|\mathbf x-\mathbf x'|}
+\frac{\delta_{L(\xi_1,\beta_1),1-L(\xi_2,\beta_2)}}{\sqrt{|\mathbf x-\mathbf x'|^2+d^2}}\right],
\end{multline}
where $d=0.335$ nm is the distance between the layers, and the layer index $L(\xi,\beta)$ is defined as
\[
L(\xi,\beta)=\left\{
\begin{array}{ll}
0,&\text{ if } \xi=1\text{ and }\beta=1,4,\\
&\text{ or }\xi=-1\text{ and }\beta=2,3,\\
1,&\text{ otherwise}.
\end{array}
\right.
\]
Substituting Eqs.~(\ref{fieldop}) and (\ref{interaction}) into Eq.~(\ref{hamiorigin}) yields the following terms:

(i) The single-particle energy difference
\[
\delta_{N\tilde N}\delta_{N'\tilde N'}\left( E_{n\alpha\xi\sigma}-E_{n'\alpha'\xi'\sigma'}\right),
\]
where $\delta_{NN'}=\delta_{\sigma\sigma'}\delta_{\xi\xi'}\delta_{nn'}\delta_{\alpha\alpha'}$.

(ii)The exchange energy difference
\begin{multline}
\Delta^{\text{ex}}_{NN'}=\int\frac{d\mathbf q}{(2\pi)^2}\sum_{M\text{ filled}}\left(\delta_{\xi'\xi_M}\delta_{\sigma'\sigma_M}I^{N'M}_{N'M}(\mathbf p)\right.\\
\left.-\delta_{\xi\xi_M}\delta_{\sigma\sigma_M}I^{MN}_{MN}(\mathbf p)\right)
\label{cost}
\end{multline}
of the promoted particle in the two bands with all particles of like spin and valley.
($I^{N_1N_2}_{N_3N_4}(\mathbf p)$ will be defined below.)

(iii) The dynamical interaction
\[
E^{(\tilde N\tilde N')}_{(NN')}(\mathbf Q)=-\int\frac{d\mathbf q}{(2\pi)^2}e^{i\mathbf{\hat z}\cdot(\mathbf q\times\mathbf Q)} I^{\tilde NN'}_{\tilde N'N}(\mathbf p).
\]

(iv) A self-energy term
\[
\Delta^{\text{RPA}(\tilde N\tilde N')}_{(NN')}(\mathbf Q)=
\frac{1}{2\pi\ell_B^2}
\mathfrak{Re}
I^{N'\tilde N}_{\tilde N'N}(\mathbf p)\\
\propto\delta_{\xi\xi'}\delta_{\sigma\sigma'}\delta_{\tilde\xi\tilde\xi'}\delta_{\tilde\sigma\tilde\sigma'}
\]
that is associated with the recombination and the recreation of the particle-hole pair, as traditionally obtained from the random phase approximation  \cite{Kallin}.
Recombination, however, is precluded in the studied problem because all excitations flip spin in state $|\text{f.m.}\rangle$ and valley in $|\text{v.p.}\rangle$.

Here
$I^{N_1N_2}_{N_3N_4}(\mathbf p)$ is the sum of a same layer contribution $S^{N_1N_2}_{N_3N_4}(\mathbf p)$ and an interlayer one $T^{N_1N_2}_{N_3N_4}(\mathbf p)$:
\begin{multline*}
S^{N_1N_2}_{N_3N_4}(\mathbf p)=\frac{2\pi e^2}{\epsilon p}\left[\right.
A^{n_1}_{\alpha_1}A^{n_2}_{\alpha_2}A^{n_3}_{\alpha_3}A^{n_4}_{\alpha_4}F_{N_1N_4}(\mathbf p)F^\ast_{N_2N_3}(\mathbf p)+\\
A^{n_1}_{\alpha_1}C^{n_2}_{\alpha_2}C^{n_3}_{\alpha_3}A^{n_4}_{\alpha_4}F_{N_1N_4}(\mathbf p)F^\ast_{N_2-1,N_3-1}(\mathbf p)+\\
C^{n_1}_{\alpha_1}A^{n_2}_{\alpha_2}A^{n_3}_{\alpha_3}C^{n_4}_{\alpha_4}F_{N_1-1,N_4-1}(\mathbf p)F^\ast_{N_2N_3}(\mathbf p)+\\
C^{n_1}_{\alpha_1}C^{n_2}_{\alpha_2}C^{n_3}_{\alpha_3}C^{n_4}_{\alpha_4}F_{N_1-1,N_4-1}(\mathbf p)F^\ast_{N_2-1,N_3-1}(\mathbf p)+\\
B^{n_1}_{\alpha_1}B^{n_2}_{\alpha_2}B^{n_3}_{\alpha_3}B^{n_4}_{\alpha_4}F_{N_1-2,N_4-2}(\mathbf p)F^\ast_{N_2-2,N_3-2}(\mathbf p)+\\
B^{n_1}_{\alpha_1}D^{n_2}_{\alpha_2}D^{n_3}_{\alpha_3}B^{n_4}_{\alpha_4}F_{N_1-2,N_4-2}(\mathbf p)F^\ast_{N_2-1,N_3-1}(\mathbf p)+\\
D^{n_1}_{\alpha_1}B^{n_2}_{\alpha_2}B^{n_3}_{\alpha_3}D^{n_4}_{\alpha_4}F_{N_1-1,N_4-1}(\mathbf p)F^\ast_{N_2-2,N_3-2}(\mathbf p)+\\
\left.
D^{n_1}_{\alpha_1}D^{n_2}_{\alpha_2}D^{n_3}_{\alpha_3}D^{n_4}_{\alpha_4}F_{N_1-1,N_4-1}(\mathbf p)F^\ast_{N_2-1,N_3-1}(\mathbf p)\right],\\
T^{N_1N_2}_{N_3N_4}(\mathbf p)=\frac{2\pi e^2}{\epsilon p}e^{-pd}\\
\left[\right.
A^{n_1}_{\alpha_1}B^{n_2}_{\alpha_2}B^{n_3}_{\alpha_3}A^{n_4}_{\alpha_4}F_{N_1N_4}(\mathbf p)F^\ast_{N_2-2,N_3-2}(\mathbf p)+\\
B^{n_1}_{\alpha_1}A^{n_2}_{\alpha_2}A^{n_3}_{\alpha_3}B^{n_4}_{\alpha_4}F_{N_1-2,N_4-2}(\mathbf p)F^\ast_{N_2N_3}(\mathbf p)+\\
A^{n_1}_{\alpha_1}D^{n_2}_{\alpha_2}D^{n_3}_{\alpha_3}A^{n_4}_{\alpha_4}F_{N_1N_4}(\mathbf p)F^\ast_{N_2-1,N_3-1}(\mathbf p)+\\
D^{n_1}_{\alpha_1}A^{n_2}_{\alpha_2}A^{n_3}_{\alpha_3}D^{n_4}_{\alpha_4}F_{N_1-1,N_4-1}(\mathbf p)F^\ast_{N_2,N_3}(\mathbf p)+\\
B^{n_1}_{\alpha_1}C^{n_2}_{\alpha_2}C^{n_3}_{\alpha_3}B^{n_4}_{\alpha_4}F_{N_1-2,N_4-2}(\mathbf p)F^\ast_{N_2-1,N_3-1}(\mathbf p)+\\
C^{n_1}_{\alpha_1}B^{n_2}_{\alpha_2}B^{n_3}_{\alpha_3}C^{n_4}_{\alpha_4}F_{N_1-1,N_4-1}(\mathbf p)F^\ast_{N_2-2,N_3-2}(\mathbf p)+\\
B^{n_1}_{\alpha_1}D^{n_2}_{\alpha_2}D^{n_3}_{\alpha_3}B^{n_4}_{\alpha_4}F_{N_1-2,N_4-2}(\mathbf p)F^\ast_{N_2-1,N_3-1}(\mathbf p)+\\
\left.
D^{n_1}_{\alpha_1}B^{n_2}_{\alpha_2}B^{n_3}_{\alpha_3}D^{n_4}_{\alpha_4}F_{N_1-1,N_4-1}(\mathbf p)F^\ast_{N_2-2,N_3-2}(\mathbf p)\right].
\end{multline*}
Here $N-1\equiv(n-1,\alpha\xi\sigma)$, and $F_{N'N}(\mathbf q)$ is related to the Fourier transform of $\eta_{nq}(\mathbf r)$ in Eq.~(\ref{etaeq}),
\begin{multline*}
F_{N'N}(\mathbf q)=\delta_{\sigma\sigma'}\delta_{\xi\xi'}\sqrt\frac{n!}{(n')!}\left(\frac{(-q_y+iq_x)\ell_B}{\sqrt2}\right)^{n'-n}\\
L_n^{n'-n}\left(\frac{q^2\ell_B^2}{2}\right)e^{-q^2\ell_B^2/4},
\end{multline*}
if $n'\ge n$, else $F_{nn'}(\mathbf q)=F^\ast_{n'n}(-\mathbf q)$.
$L^m_n(z)$ is an associated Laguerre polynomial.
Notice that $E^{(\tilde N\tilde N')}_{(NN')}(\mathbf Q)\propto\delta_{\tilde n'-\tilde n,n'-n}$ is not necessarily diagonal as a matrix indexed by Landau level pairs.

\section{The exchange energy cost of the $l_z=\pm1$ excitations}

The transitions from $n=0$ to $(n=1,\alpha=2)$ orbitals involve a large energy cost 
due to exchange with an infinity of filled Landau levels.
If an electron is promoted from an $n=0$ orbital to an $n=1$ Landau orbital in either the ferromagnetic state $|\text{f.m.}\rangle$ or the valley-polarized one $|\text{v.p.}\rangle$ ($l_z=1$ excitation),
the exchange energy with an infinite number of completely filled Landau bands ($n\ge2$, $\alpha=2,3$, all combinations of $\xi=\pm1$ and $\sigma=\uparrow,\downarrow$) contributes to the energy shift $\Delta^{\text{ex}}_{(1,\alpha=2,\xi\sigma)(0\xi'\sigma')}$;
an analogous statement holds for the $l_z=-1$ transition.
This number may be formally infinite because we pick up a contribution from very low energies where the Hamiltonian $\hat H_\xi$ in Eq.~(\ref{hamilton}) is no longer valid.
This calls for some renormalization procedure.
Exploiting the particle-hole symmetry of undoped bilayer graphene, however, this turns out to be a very simple one.
Consider the ferromagnetic state $|\text{f.m.}\rangle$ for concreteness.
The exchange costs of the $l_z=1$ and $l_z=-1$ transitions are
\begin{align*}
\Delta^{\text{ex}}_{(1,\alpha=2,\xi\sigma)(0\xi'\sigma')}&=\Delta^{\text{ex}}_\infty+\int\frac{d\mathbf q}{(2\pi)^2}I^{00}_{00}(\mathbf p),\\
\Delta^{\text{ex}}_{(0\xi\sigma)(1,\alpha=2,\xi'\sigma')}&=-\Delta^{\text{ex}}_\infty+\int\frac{d\mathbf q}{(2\pi)^2}I^{1\alpha=2,1\alpha=2}_{1\alpha=2,1\alpha=2}(\mathbf p),
\end{align*}
respectively.
As the state obtained by moving an electron from the $(0\xi\downarrow)$ band to the $(1,\alpha=2,\xi\uparrow)$ band is the particle-hole conjugate of the state that is obtained 
by moving a hole from the  $(1,\alpha=2,\xi\uparrow)$ band to the $(0\xi\downarrow)$ band, and the number of identical spin and identical valley bands is the same in the two cases,
\[
\Delta^{\text{ex}}_\infty+\int\frac{d\mathbf q}{(2\pi)^2}I^{00}_{00}(\mathbf p)=-\Delta^{\text{ex}}_\infty+\int\frac{d\mathbf q}{(2\pi)^2}I^{1\alpha=2,1\alpha=2}_{1\alpha=2,1\alpha=2}(\mathbf p).
\]
In the special case of the two band model this consideration yields $\Delta^{\text{ex}}_\infty=-\frac{\sqrt{2\pi}}{8}\frac{e^2}{4\pi\epsilon_0\epsilon_r\ell_B}$;
here the infinite summation  in $\Delta^{\text{ex}}_\infty$ can actually be performed and yields an identical result.
In the four-band model, on the other hand, the corresponding value ranges decreases from $\sim\frac{-0.16e^2}{4\pi\epsilon_0\epsilon_r\ell_B}$ to $\sim\frac{-0.21e^2}{4\pi\epsilon_0\epsilon_r\ell_B}$
in the interval from $B=1$ to $B=40$ T, and shows a tiny interlayer energy $\Delta_L$ and valley $\xi$ dependence.

\newcommand{\PRL}{Phys.\ Rev.\ Lett.}
\newcommand{\PRB}{Phys.\ Rev.\ B}
\newcommand{\PR}{Phys.\ Rev.}
\newcommand{\NPB}{Nucl.\ Phys.\ B}


\begin{thebibliography}{99}
\bibitem{fnote} Incidentally, this sketch was drawn using the two-band model,\cite{tightbinding} which applies for small momenta ($\ll\gamma_1/\hbar v$) and small energies ($\ll\gamma_1$).
\end{thebibliography}
\end{document}